# A Neural Vocoder Based Packet Loss Concealment Algorithm

*Yao Zhou, Changchun Bao\**

Speech and Audio Signal Processing Laboratory, Faculty of Information Technology,
Beijing University of Technology, Beijing, China
zhouyao@emails.bjut.edu.cn, baochch@bjut.edu.cn

## Abstract

The packet loss problem seriously affects the quality of service in Voice over IP (VoIP) sceneries. In this paper, we investigated online receiver-based packet loss concealment which is much more portable and applicable. For ensuring the speech naturalness, rather than directly processing time-domain waveforms or separately reconstructing amplitudes and phases in frequency domain, a flow-based neural vocoder is adopted to generate the substitution waveform of lost packet from Mel-spectrogram which is generated from history contents by a well-designed neural predictor. Furthermore, a waveform similarity-based smoothing post-process is created to mitigate the discontinuity of speech and avoid the artifacts. The experimental results show the outstanding performance of the proposed method.

## 1. Introduction

Nowadays, voice over Internet Protocol (VoIP) has become increasingly popular. To address the unreliable delivery of packets over the Internet and guarantee the quality of service (QoS) [1], many packet loss concealment (PLC) methods have been developed and refined.

This paper is focused on the reformation of receiver-based PLC algorism [2]. Since the human hearing has a masking effect and the speech signal has short-term self-similarity, the receiver-based PLC techniques attempt to partially recover the speech signal of a lost packet only from its history or future information.

According to [3], the traditional receiver-based PLC schemes are categorized into insertion-based, interpolation-based, and regeneration-based approaches. Apart from trivial methods like silence substitution, the waveform similarity overlap-and-add (WSOLA) method [4] and hidden Markov model (HMM)-based method [5] are more intelligent. Nevertheless, these methods suffer from the artifacts and may cause catastrophic clipped-speech distortion, especially when dealing with long gaps and transients.

Recently, deep learning has been introduced to the PLC. According to [6], deep learning-based PLC may be classified into offline and online systems. The offline PLC system estimates lost packets using a large chunk of audio including the lost parts as the input, such as generative adversarial network (GAN) based frameworks [7,8,9] and auto-encoder based frameworks [10,11]. However, the latency introduced by these systems is absolutely unacceptable.

Correspondingly, the online PLC system predicts lost packets in real time, which only requires history information of speech signal. In [12], two deep neural networks (DNN) were separately trained to predict the log-power spectra and phases of the lost packets. As a feasible way to get rid of the difficulty of phase prediction, recurrent neural network (RNN) was adopted for PLC tasks in time domain [6, 13]. In addition, a non-autoregressive adversarial auto-encoder was also proposed to perform real-time PLC in time domain [14]. In [15], the frequency-domain spectrum and time-domain waveform were combined as the input of an encoder-based framework. The WaveNetEQ [16] using WaveRNN was proposed to recover the signal of the last frame from a log-Mel spectrogram of previous frame.

To our best knowledge, the performance of time-domain method may be limited, because of the high temporal resolution of waveform samples. And phase reconstruction hinders the spectra-domain method. The time-domain waveform synthesized by most works mentioned above lack naturalness and continuity in the listening.

For improving the naturalness, inspired by the achievement in Text-to-Speech Synthesis (TTS), rather than constructing candidate phase for the lost speech, a generative model was sought to synthesize speech waveform from time-frequency (T-F) representations. The state-of-the-art WaveNet [17] used in Tacotron2 [18] could generate highly naturalness waveform from spectrum of frequency-domain, but the time complexity is unacceptable. In [19], a flow-based model, named Waveglow, is well suited to today's massively parallel computers and simple to train, in which its naturalness has been shown a rival of human voice.

In this paper, we propose a novel deep learning-based online framework for PLC. The framework consists of a deep neural network (DNN), a flow-based vocoder and a smooth catenation process. The effectiveness of the DNN has been verified in log-power spectrum prediction in [12], so it is adopted in this work to predict Mel-spectrum of the lost packet from previous valid packets. The predicted Mel-spectrum is transformed into time-domain waveforms by a neural vocoder. A smoothing post-process is used to solve the problem of waveform discontinuity. In order to demonstrate the superiority of the proposed algorithm, we compare it to several neural PLC frameworks and classical WSOLA. According to objective tests and waveform visualization, it turns out that the proposed PLC algorithm outperforms some existing algorithms.

The remainder of this paper is organized as follows: after briefly introducing the packet loss simulator in section 2. Section 3 describes the architecture of the proposed PLC algorithm. Then, experiments are presented in Section 4. At last, Section 5 summarizes our contributions.

## 2. Packet Loss Simulator

To create training and test data for PLC, it is essential to simulate realistic scenarios of packet loss. According to [21], Both Bernoulli model and two-state Markov model are not appropriate for many real-world scenarios.

In this paper, following the ITU standard [22], the Gilbert-Elliot model shown in Figure 1 is adopted. As shown in Figure 1, "G" denotes good channel state and "B" denotes bad channel state. The parameters $\alpha$ and $\beta$ are the transition probabilities from state G to stated B and from state B to stated G, respectively. $P_G$ and $P_B$ are the packet loss probabilities in states G and B, respectively.

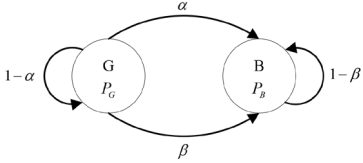

Figure 1: *Gilbert-Elliot model*

The mean packet loss rate generated by this model is given by

$$PLR = \frac{\alpha}{1-\lambda} * P_B + \frac{\beta}{1-\lambda} * P_G \quad (1)$$

where $\lambda = 1 - (\alpha + \beta)$ is an indication of the burst or random characteristics of the channel. In this issue, usually $\lambda > 0$, and if $\lambda = 0$, the model is reduced to the Bernoulli model. Once the packet loss rate (*PLR*) is given, which is reasonable only in the range $0 \leq PLR \leq 0.5$, the transition probabilities could be derived as

$$\alpha = (1-\lambda)*(1-\frac{P_B - PLR}{P_B - P_G}) \quad (2)$$

and

$$\beta = (1-\lambda)*\frac{P_B - PLR}{P_B - P_G} \quad (3)$$

## 3. Proposed Method

The flowchart of the proposed PLC algorithm is shown in Figure 2. In the inference stage, any packet is considered an input whether the packet is lost or not. If a packet is not lost, it is directly copied to the output and loaded into a memory buffer. Otherwise, the previous packets drawn from the memory buffer are used as the inputs of neural PLC framework to predict the lost packet.

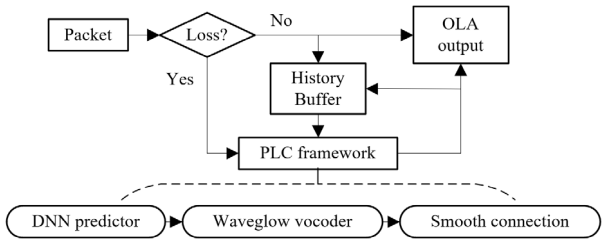

Figure 2: *Flowchart of the proposed PLC algorithm.*

We found that when continuous packets are lost, an attenuation factor is unnecessary in many existing PLC methods. So, in this paper, the predicted packets are treated as the received ones. For clarity, in this paper, one packet includes the information of one frame.

### 3.1. DNN Predictor

The receiver-based PLC is usually applied as a post-processing step after the decoding operation and does not affect bit-stream compatibility. Since the feature parameters in the lost packets are unavailable, the predictor of the feature parameters used for recovering time gaps become particularly necessary. In this work, a fine designed deep neural network (DNN) is applied as the predictor.

First of all, an appropriate feature representation must be chosen for successful prediction. Although the log-power spectrum in the Short-time Fourier transform (STFT) domain is a very suitable feature, its high feature dimension could be further reduced by transforming it into Mel-spectrum. And also, the Mel-spectrum has been widely employed in many natural language processing (NLP) tasks [23, 24]. Based on this, the Mel-spectrum is selected as the feature in this paper.

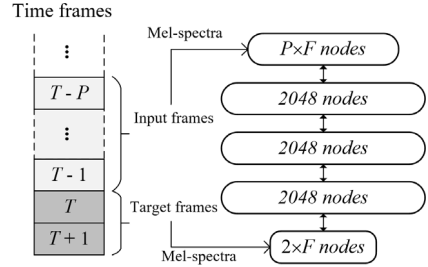

Figure 3: *Structure of the proposed DNN predictor*

As shown in Figure 3, in the training stage, the waveform sequences $\mathbf{x}_L$ with the length $L$ in time-domain are segmented into the frames by a window overlapped, where the frame length is $W$ and frame shift is $H$.

In which case, a Mel-spectrum vector of current frame is denoted as $\mathbf{m}_T$ consisting of $F$ Mel-bands. The input layer with $P \times F$ nodes consists of previous $P$ frames, while $\mathbf{m}_T$ and $\mathbf{m}_{T+1}$ are concatenated to be the target features. The nodes of three hidden layers are all set as 2048. A logistic sigmoid function is used for the activation function of the hidden units and the type of the output layer with $2 \times F$ nodes is linear. Also, the features are normalized to zero mean and unit variance. The Minimum Mean Square Error (MMSE) between estimated Mel-spectra and the reference target Mel-spectra is conducted as the objective criterion. The inference process is consistent with the forward propagation in the training.

### 3.2. Flow-based Vocoder

After obtaining T-F representations, a vocoder is needed to transform them into the waveforms. Traditional methods like Griffin-Lim algorithm (GLA) [25] and PGHI [26] are not suitable for PLC tasks. The flow-based neural generator is first introduced in [20]. And later, Waveglow [19] was proved that the flow-based model is capable of generating high quality speech from Mel-spectrogram. The main purpose of this architecture is to use a neural model to realize an invertible transformation $f$ between a simple distribution $\mathbf{z}$ and the desired distribution of speech samples $\mathbf{x}$ conditioned on a Mel-spectrogram, i.e.,

$$\mathbf{z} = f(\mathbf{x}), \quad \mathbf{x} = f^{-1}(\mathbf{z}) \quad (4)$$

where $f(\cdot)$ (and likewise, $f^{-1}(\cdot)$) is composed of a sequence of invertible transformations.

$$f = f_0 \circ f_1 \circ ... f_{k-1} \circ f_k \quad (5)$$

This model is directly trained to minimize the negative log-likelihood of the data, which is formed by (6), where $\mathbf{J}$ in (6) denotes Jacobian matrix.

$$\log p_\theta(\mathbf{x}) = \log p_\varepsilon(\mathbf{z}) + \sum_{i=0}^{k} \log\left|\det\left(\mathbf{J}(f_i(\mathbf{x}))\right)\right| \quad (6)$$

The structure of our vocoder is similar to the Waveglow. The invertibility is achieved by a series of affine coupling layers described by the following equations (7) ~ (10) and invertible $1 \times 1$ convolution layers described by equation (11) [19, 27].

$$[\mathbf{x}_0, \mathbf{x}_1] = \mathbf{x} \quad (7)$$

$$\begin{cases} \boldsymbol{\delta}_0 = \mathbf{x}_0 \\ \boldsymbol{\delta}_1 = \mathbf{s}_i(\mathbf{x}_0) \otimes x_1 + \mathbf{t}_i(\mathbf{x}_0) \end{cases} \quad (8)$$

$$(\log(\mathbf{s}_i), \mathbf{t}_i) = WN(\mathbf{x}_0, \text{Mel-spectrogram}) \quad (9)$$

$$f_i^{coupling}(\mathbf{x}) = [\boldsymbol{\delta}_0, \boldsymbol{\delta}_1] \quad (10)$$

$$f_i^{conv}\left(f_i^{coupling}(\mathbf{x})\right) = \mathbf{W}_i \bullet f_i^{coupling}(\mathbf{x}) \quad (11)$$

The $WN(\cdot)$ in (9) represents a residual network which is similar to WaveNet [17] that is no causal convolution. Finally, the negative likelihood is given by

$$\log(p_\theta(\mathbf{x})) = \log(p_\varepsilon(f(\mathbf{x}))) + \sum_{i=0}^{k}(\log \mathbf{s}_i) + \sum_{i=0}^{k}(\log \det|\mathbf{W}_i|) \quad (12)$$

In this work, to further improve the speed of speech synthesis, the affine coupling layers and invertible convolutions are truncated to 10 from 12. While training, the speech samples and corresponding Mel-spectrograms are used as the model input. The frame length, shift length and the length of speech samples in each iteration are adjusted to satisfy the PLC tasks. Figure 4 illustrates the inference process of time-domain speech generation. The model mentioned above is inversed to be a vocoder to generate the speech waveforms. To better address the correlation of speech signal, rather than only use two predicted frames, the Mel-spectrograms of previous $P$ frames are catenated with the two predicted Mel-spectrograms. Thus, totally, the Mel-spectrograms of ($P$+2) frames are feed into the vocoder as condition constraints. The length of waveform corresponding to ($P$+2) frames is set as $L$. Then, by randomly collecting data $\mathbf{z}_L$ from Gaussian distribution until the length of the data reaches to $L$, speech waveform $\tilde{\mathbf{x}}_L$ with the length $L$ would be generated by simply running Mel-spectrogram and $\mathbf{z}_L$ through the vocoder.

### 3.3. Smooth Connection

Because the flow-based vocoder pursues statistical similarity of real speech simples, if the substitution of the lost frame is directly chosen from the corresponding frame in the generated waveform $\tilde{\mathbf{x}}_L$, the discontinuity of the waveform will occur after overlapping and adding. This seriously degrades the naturalness of speech in the listening.

The way to solve this problem is to find a frame that is most close to the adjacent one. A method based on waveform similarity is proposed in this paper for finding the proper substitution. Figure 5 and Figure 6 show an example of this operation.

In Figure 5, there is an overlapped part between the frames. The part of most currently previous frame which is overlapped with the lost frame is selected as a pattern $A(n)$, where $n$ denotes discrete time index. Then, the last three frames of waveform $B(n)$ generated by the vocoder are sent to calculate the cross-correlation with pattern $A(n)$. The discrete time index corresponding to the maximum cross-correlation is set as a start point for truncating a segment with the length $W$ as the substitution of the lost frame, as depicted in Figure 6. Finally, this frame is outputted and sent into the history buffer.

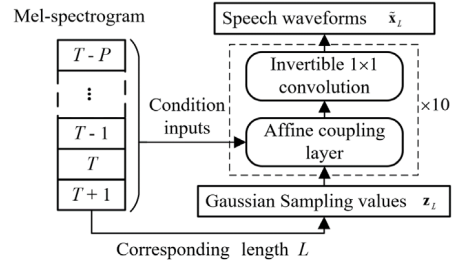

Figure 4. *Inference flowchart of the flow-based vocoder.*

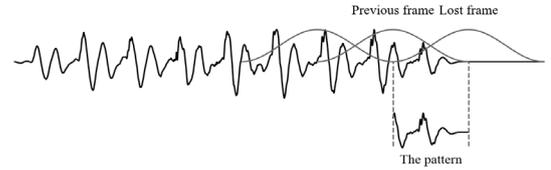

Figure 5. *Selection of the overlapped part*

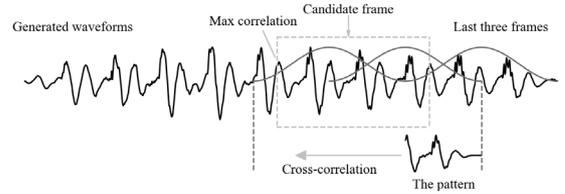

Figure 6. *Determination of substitution of the lost frame*

## 4. Experiments

### 4.1. Settings

For all experiments, the LJ speech corpus sampled at 16 kHz [28] in a home environment is adopted. This corpus consists of 13100 short audio clips of a single speaker and the length of each clip varies from 1 to 10 seconds. The total length of this corpus is about 24 hours.

The Mel-spectrogram is obtained by a 512-points FFT on each frame with the length $W = 320$ samples (20ms), shift length $H$ of the frame is 160 samples (10ms). The number of Mel filters are 80 ($F$=80). For DNN predictor, the input contains 11 previous frames (120ms), i.e., $P = 11$. For packet loss simulator, $\lambda = 0.5$, $P_G = 0$, which means error free in state "G", and $P_B = 0.5$ in state "B". Thus, a totally uncertain loss happens in state "B". And one packet is assumed to contain only one speech frame. Four kinds of packet loss rate ( $PLRs$ vary from 10% to 50%) are simulated and tested. The DNN predictor and flow-based vocoder are trained separately at first, and then they are trained together.

### 4.2. Evaluation

The perceptual evaluation of speech quality (PESQ) [29], short-time objective intelligibility (STOI) [30] and log-spectrum distortion (LSD) [31] are conducted as the objective measures to evaluate the speech quality with the PLC. Because there are not enough experts to accomplish formal subjective listening tests, we uploaded some demos at website: https://github.com/AugggRush/PLC-demos.

In this evaluation, the silence substitution scheme is regarded as the baseline. The widely used WSOLA [4] method is chosen as a traditional comparation. The DNN [12] integrated with GLA method (DNN&GLA) and recent encoder-based framework (Encoder) [15] are selected as neural comparation.

Table 1: *Test results of PESQ, STOI and LSD*

|  | PLR | 10% | 20% | 30% | 50% |
|---|---|---|---|---|---|
| PESQ | Silence | 2.62 | 2.24 | 1.80 | 1.07 |
|  | DNN&GLA | 2.66 | 2.31 | 1.84 | 1.51 |
|  | WSOLA | 3.02 | 2.58 | 2.28 | 1.69 |
|  | Encoder | 3.20 | 2.77 | 2.48 | 1.92 |
|  | Proposed | **3.32** | **2.89** | **2.60** | **2.07** |
| STOI | Silence | 0.942 | 0.890 | 0.834 | 0.715 |
|  | DNN&GLA | 0.946 | 0.908 | 0.870 | 0.795 |
|  | WSOLA | 0.948 | 0.911 | 0.869 | 0.790 |
|  | Encoder | 0.974 | 0.948 | 0.918 | 0.842 |
|  | Proposed | **0.980** | **0.958** | **0.935** | **0.873** |
| LSD | Silence | 1.28 | 2.69 | 4.78 | 9.14 |
|  | DNN&GLA | 0.77 | 1.46 | 2.09 | 3.05 |
|  | WSOLA | 0.62 | 1.14 | 2.57 | 3.56 |
|  | Encoder | 0.48 | 1.02 | 1.79 | 3.54 |
|  | Proposed | **0.35** | **0.72** | **1.11** | **2.06** |

Table 1 describes the wide-band PESQ, STOI and LSD scores in various packet loss rates, respectively. Compared with the baseline, DNN method integrated with GLA performs much worse than satisfactorily, there is only slight improvement at high *PLR*. This confirms that GLA is not a good vocoder for PLC tasks. The WSOLA method derives better PESQ than DNN&GLA, but its STOI is lower at 30% and 50% loss rates. This might because long gaps at high loss rates contain new phonemes, that is, it stretches the waveform too long to restore the correct phonemes and results in a loss of intelligibility. Encoder method performs much higher than DNN&GLA on PESQ and STOI. Besides, it should be noticed that both WSOLA and Encoder are worse than DNN&GLA in LSD scores, because DNN only deal with the power spectrum which is more structural, while the other two methods consider amplitude and phase spectra together.

Furthermore, in all objective tests, the proposed method is far better than all reference methods in various *PLR*s. Compared to the baseline, our method averagely increases the PESQ and STOI remarkably by 46% and 11.5%. With the increase of the PLR, the quality improvement is also increased, because more fragments need to be repaired while the loss becomes large. It is notable that our method almost increases 1 PESQ score at 50% loss rate, and except for 50%, the proposed method obtains nearly transparent LSD scores.

Figure 7 shows some examples of waveform transition between unvoiced and voiced speech. In the first grey box, it is obvious that the proposed vocoder method recovers transition from the voiced to unvoiced waveform far better than other methods. For the transition from the unvoiced to voiced waveforms in second and third box, all methods fail to recover the energy of the voiced speech from the low-energy unvoiced speech.

Figure 8 shows an example of the voiced waveforms. The first box contains a single frame loss (10ms), meanwhile, the second box contains a consecutive loss of five frames (50ms). The waveform recovered by DNN&GLA is the worst that is away from the origin one. The WSOLA method can recover the pitch, but it suffers from severe energy loss. And both Encoder and proposed methods track the pitch trajectory very well, but as the continuous packet loss increases, the energy of recovered waveforms become smaller and smaller. In comparation, the proposed method obtains a result closer to origin one and enriches much more waveform details, which makes the speech sound more real and natural. This might be caused by the difference between the training targets of two neural models. The flow-based vocoder tries to learn probability distribution of the speech, and then generates waveforms by sampling from the learned distribution, which means that any value has a greater or lesser probability of being taken. However, the Encoder method trains a CNN model to minimize the mean square error (MSE) between the generation and target, which makes the generation like a linear approximation of real waveform.

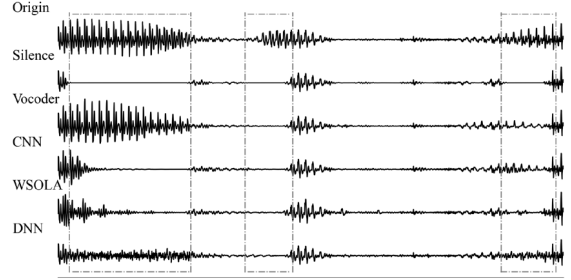

Figure 7. *Comparison of waveform transition*

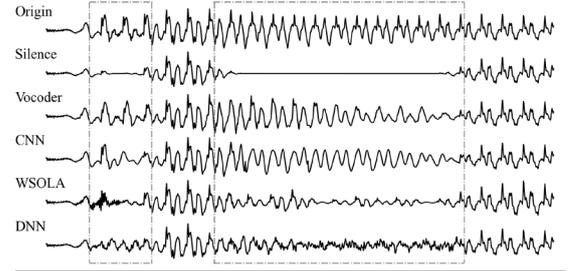

Figure 8. *Voiced waveform comparison*

## 5. Conclusions

In this work, we proposed to use a neural architecture to accomplish an online PLC task. A DNN is designed to predict the Mel-spectrograms of the lost packets from the history contexts only. A flow-based vocoder was used to transform the Mel-spectrograms into natural speech waveform. Waveform similarity-based smoothing post-process was utilized to find the most similar substitution to mitigate the discontinuity. Compared with reference methods, the proposed method absolutely has a better performance. In the further work, a formal subjective test can be conducted to further verify the advantage of our system. For more accurately predicting Mel-spectrograms, a more effective predictor could be exploited, and also it is necessary to find a way to solve the energy decay problem in consecutive packet loss.

## 6. Acknowledgment

This work was supported by the National Natural Science Foundation of China (Grant No.61831019).